# 2D wavelength-polarization dispersive microspectroscope based on a hybrid plasmonic helical nanostructure

*Zhiguang Sun*[a], *Huan Chen*[b], *Zhenglong Zhang*[b], *Lujun Pan*[a], *Yiming Yang*[c], *Bin Dong*[d,*] *and Yurui Fang*[a,*]

[a] Key Laboratory of Materials Modification by Laser, Electron, and Ion Beams (Ministry of Education); School of Physics, Dalian University of Technology, Dalian 116024, P.R. China

[b] School of Physics and Information Technology, Shaanxi Normal University, Xi'an 710062, P. R. China

[c] School of Microelectronics, Dalian University of Technology, Dalian 116024, P.R. China

[d] School of Physics and Materials Engineering, Dalian Minzu University, Dalian 116600, P. R. China

[*] Corresponding authors: yrfang@dlut.edu.cn (Fang); dong@dnu.edu.cn (Dong)

# Abstract

Microspectrometer features remarkable portability and integration, and is prospected to provide quick detection tools to daily life and revolutionary techniques to researchers. For the extremely finite footprint, microspectrometer can hardly work to analyze the polarization feature by placing polarizer in the optical path like conventional spectrometers. Here, we demonstrate a novel 2D wavelength-polarization dispersive microspectroscope based on carbon nanocoil with plasmonic Au nanopariticles (Au/CNC) as a dispersive component. Explored by the microspectrum and Fourier-space microscopy, a unique 2D dispersive characteristic of the Au/CNC is revealed. Along the axis of the coil, Au/CNC disperses light as wavelength to bands of different diffraction orders like a grating. Wavelength of the incident light can be obtained from the position of the signal lines in a quite large visible-near-infrared wavelength range with an acceptable resolution. In the direction perpendicular to the axis of the coil, incident light is dispersed as polarization with bright and dark areas. We can distinguish left- and right-circularly-polarized light, and also obtain the polarization orientation of linearly-polarized light. Based on this effect, a wonderful 2D wavelength-polarization microspectrometer can be built. It not only fulfills the wavelength analysis with a tiny dispersive component, but also simultaneously knows the polarization feature of the incident light in one shot. Moreover, this powerful tool can further evolve new revolutionary techniques via integrated with other systems.

# Keywords

Microspectrometer; localized surface plasmon; Fourier image; light splitting; polarization

# Introduction

Spectrometer is one of the most important instruments in the laboratory for convenient and nondestructive material characterization and chemical analysis. However, most of the high-performance spectrometers can only work by transporting samples to the laboratory, and the large footprint and high cost severely district its applications [1]. In many scenarios, such as outdoor crop and animal monitoring, environment analysis, incident investigation, it is just needed to detect signature spectral peaks instead of accurate quantitative analysis, and fast deployment of the instruments with on-the-spot results are in demand [2]. The miniaturized spectrometers with a relatively low resolution (in the order of 10 nm) and good portability well meet the requirements above. Further minimize the spectrometer to microscale, the remarkable portability and integration broaden the application of the powerful tool in new horizons. For instance, microspectrometer can be integrated into smartphones. Combining the artificial intelligence, everybody can conveniently analyze the food, cosmetics, any other daily items, and also the skin, metabolites and body fluids, thus monitor our health and even develop applications beyond imagination, which will greatly improve our lives [3]. On the other hand, microspectrometers are easily integrated into other systems, and thus bring revolutionary new techniques, such as lab-on-chip spectroscopy, in vitro characterization, spectrometer-per-pixel snapshot hyperspectral imaging, etc. [1, 4].

After years of investigation, microspectrometers based on different strategies have been developed, such as the ones splitting light spacially [5-7], filtering light by tunable or arrayed narrowband filters [8-10], Fourier transforming with light interference [11-13], and computationally reconstructing spectra [14-16]. Dispersive optics-based spectrometers that splitting light spacially to detectors have the similar configuration with conventional spectrometers, obtain extensive exploration and applications [1, 17]. The minimization of the dispersive component is essential to minimize the spectrometer system. Benefiting from the modern micro/nanofabrication techniques, microgratings with designed structures for light splitting can be well fabricated [17-19]. However, as the miniaturization goes further, quality issues of the component emerge, such as light scattering on surfaces with etching-induced roughness [1, 20]. Besides, the costly facilitates raise the investigation and application threshold.

Polarization is the other key property of light besides wavelength and intensity. It tells about the surface features of the objects and chiral nature of structures and molecules, playing an important role in the investigation of materials, nanophotonics, biology, pharmacy, astronomy, etc. [21-24]. Comprehensively analyzing the polarization, wavelength and intensity of light, deep information in different aspects can be revealed. Since the discovery of birefringence in crystals, polarimetric measurements are still mainly conducted by placing polarizers in the optical path. The wavelength or space distribution of the light with polarization condition can be obtained by collecting signals at different polarizer axis orientations [22-23]. The low collection efficiency is a serious problem of this method, for the light is repeatedly collected at every polarizer axis orientation. In the cases of quite low light intensity (long integrated time is needed) or high polarimetric resolution, the extremely long collection time is unacceptable. This also means the conventional method is difficult to analyze the dynamic objects with weak signals. For microspectrometer, the extremely finite footprint inhibits the addition of polarizer, thus the present microspectrometers are incompetent for polarization analysis. Therefore, the development of new spectrometry techniques and microspectrometers with wavelength-polarization simultaneous analysis ability is of great significance and will bring about a profound revolution in various fields.

Carbon nanocoil (CNC) is a kind of carbon materials with great 3D helical structure. It presents a variety of excellent properties, such as super-elasticity, large area, tunable resistance, electromagnetic wave absorption, field emission, etc., and attracts extensive attention and investigation [25-28]. Here, we employed CNC supported with Au nanoparticles (Au/CNC) as dispersive component for microspectroscope, based on the grating-like periodic structure of the CNC and the strong light-matter interaction enhanced by the localized surface plasmon resonance (LSPR) of Au nanoparticles (NPs) [29]. Analyzed by Fourier plane imaging with the microscope, the Au/CNC exhibits an amazing 2D light dispersion phenomenon. In the direction along the axis of Au/CNC, the light is dispersed to bands of different diffraction orders like a grating, while in the direction perpendicular to the axis of Au/CNC, the band areas of bright and dark vary with the polarization condition of the incident light. Based on this effect, a wonderful 2D wavelength-polarization microspectrometer can be built.

## Results and discusstion

**Characterization of the Au/CNC**

Au/CNC was prepared by a two-step synthesis route as shown in Figure 1a : (i) chemical vapor deposition (CVD) of CNC on the $\alpha$-$Fe_2O_3$/$SnO_2$ catalyst with acetylene; (ii) direct growth of Au NPs via sodium citrate reduction of chloroauric acid on the dispersed CNCs in the solution. The appearance of the as-prepared Au/CNC was shown in SEM images of Figure 1b, c. It is clearly that the CNC has a great periodic helical structure, and NPs with diameter about 20 nm uniformly disperse on the surface of CNC. The heterogeneous structure was further analyzed by high-angle annular darkfield scanning transmission electron microscopy (HAADF-STEM) and energy-dispersive X-ray (EDX) elemental mapping. The image and corresponding Au, C mapping in Figure 1d, e, f well indicate the attribution of bright NPs and dark helical region to Au and C, respectively. The Au NPs can also be revealed by the high-resolution TEM (HRTEM) image in of the region of yellow square, respectively. (h) SEAD pattern of the Au/CNC. The scale bar in Figure 1b, c, d, g and the inset of Figure 1g respectively indicate 2 μm, 500 nm, 200 nm, 10 nm and 2 nm.

Figure 1g. The NP is polycrystal, and the lattice fringes well correspond with that of Au (JCPDS #04-0784). The crystal orientation of the Au NPs was further illustrated by the fast Fourier transform (FFT) image in the inset of Figure 1g and the selected-area electron diffraction (SAED) pattern in Figure 1h.

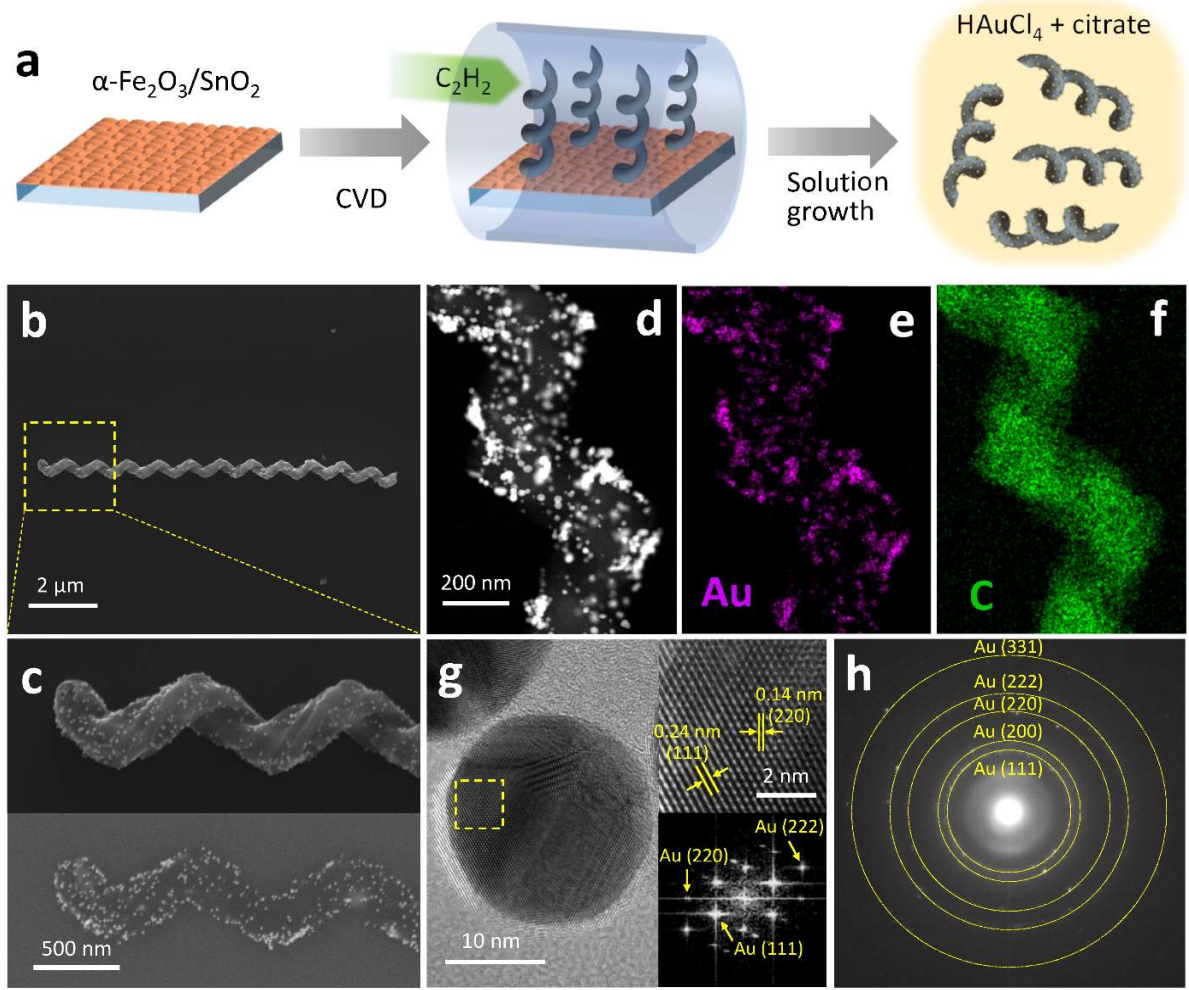

**Figure 1.** (a) Schematic illustration of the two-step route to prepare Au/CNC. (b, c) SEM images of the Au/CNC. The images in the top and bottom panels of Figure 1c were respectively collected by secondary electron and back-scattered electron detectors. (d) HAADF-STEM image and corresponding EDX elemental mapping of (e) Au and (f) C for the Au/CNC. (g) HRTEM image of Au nanoparticles on Au/CNC, the top and bottom insets are the enlarged image and FFT pattern

**Optical response of the heical structure**

Au NPs feature strong LSPR in the visible range. We growed Au NPs on the surface of CNC to improve the optical response of the structure, and hence improve its sensitivity for optical devices. Single Au/CNC and CNC with similar diameter, pitch and effective length (blocking CNC longer part) were explored and compared. Their reflective-dark-field spectra in Figure 2a show that the primary scattering peak of the helical structure red shifts from approximate 490 nm to 570 nm after the addition of Au NPs, with the green scattered light turning into yellow in the insets. Benefiting from LSPR of the Au NPs, Au/CNC has stronger interaction with light than CNC, and it has a significant scattering enhancement (20 - 60%) at 550-750 nm, with a small scattering loss (<15%) at 410-500 nm due to the scattering surface reduction of CNC for the covering of Au NPs. Besides the enhancement of scattering, the Au NPs also modify the scattering direction of the helical structure, which will be presented in the following monochromatic analysis part. The dark-field spectra with the polarization analyzer angle of 0° (P0) and 90° (P90) were shown in Figure 2b. Similarly, Au/CNC has stronger scattering intensity than CNC at both emission polarization angles, and those of P90 are stronger than those of P0. Two primary peaks around 500 and 750 nm with secondary peak oscillations about 112 nm period can be clearly observed for Au/CNC. The primary peaks are attributed to the transverse and longitudinal excitation of the long coil structure, and the secondary peak oscillations are attributed to the interference of the light emission from the periodic coil curves.

The helical Au/CNC has a great chiral structure, and promises to have different optical response to left- and right-circularly-polarized (LCP, RCP) light. With this in mind, the side-incident circularly-polarized illumination experiments were performed. The schematic setup and the definition of polarizer angles and coil orientation are illustrated in Figure 2c. The spectra in Figure 2d1 show two peaks at similar wavelengths with those in Figure 2b. The transverse incident condition (S = 90°) has spectra intensity folds of that at longitudinal incident condition (S = 0°) due to the larger scattering cross section at the former condition. The chiral structure enables the coil to scatter more for RCP light, and the circular dichroism (CD = $\frac{\sqrt{I_R}-\sqrt{I_L}}{\sqrt{I_R}+\sqrt{I_L}}$, $I_R$, $I_L$ are the spectra intensity for RCP and LCP incident conditions, respectively) in Figure 2d2 shows quite large values (0.2-0.3). For the polarization dependent spectra in Figure 2e1, f1, both intensity and profile of the spectra for RCP and LCP incident light exhibits significant difference, especially for the condition of P90. The main difference of the spectra profile for the different circularly-polarized light appears at about 700-900 nm, and it is the longitudinal mode of the coil. For both longitudinal and transverse illumination, the CD at P90 (0.4-0.6, Figure 2e2, f2) is even approximately twice of that without analyzer (0.2-0.3) in visible range, indicating a significantly difference of the polarized emission light for RCP and LCP incident conditions. Besides, it is very interesting that the emission light with polarization orientation along the axis of the coil (P90) is almost absent for LCP incident condition in a very large wavelength range of 420-800 nm. It implies the scattered light of the LCP incident light polarized almost perpendicular to the axis of the coil.

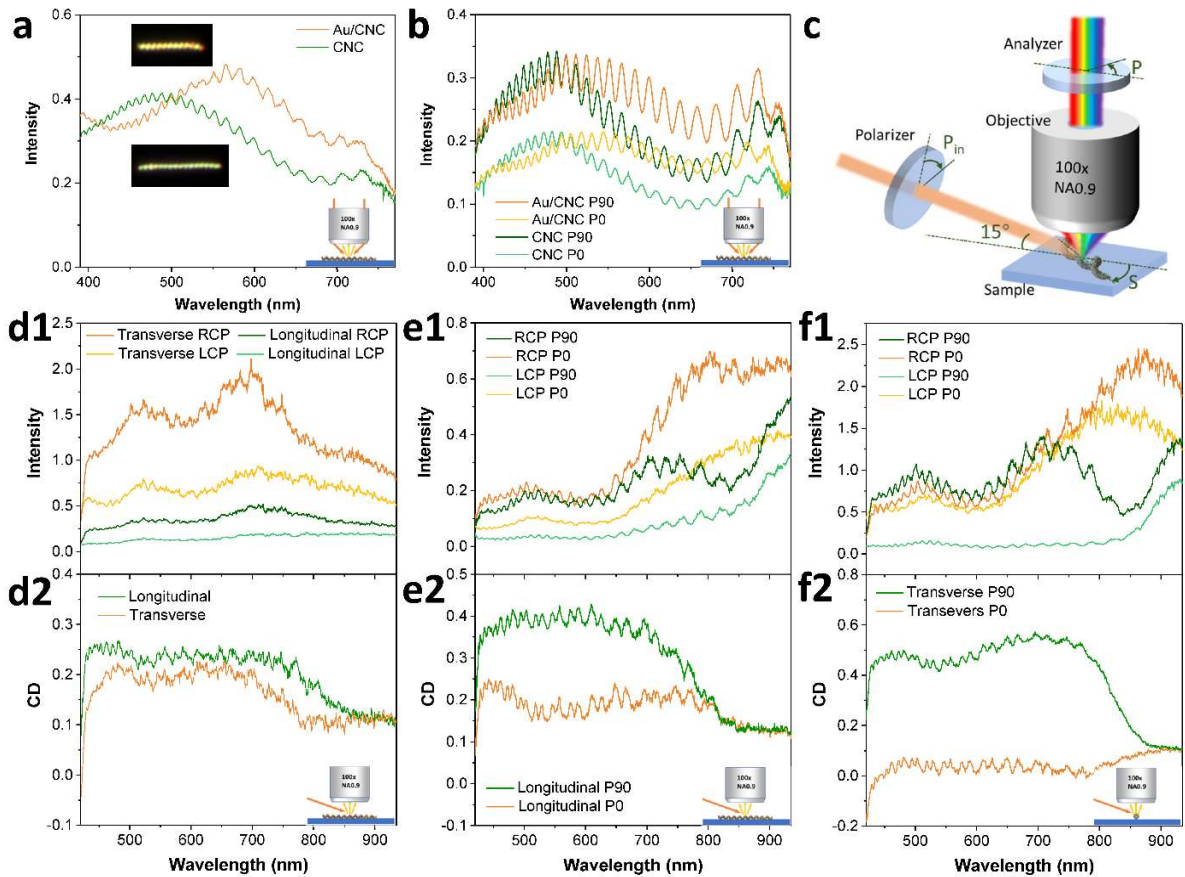

**Figure 2.** (a) Dark-field spectra of Au/CNC and CNC with light source of the microscope. The insets are the dark-field images of Au/CNC and CNC. (b) Dark-field spectra of Au/CNC and CNC with polarization analyzer angles of 0° and 90° to the coils. (c) Schematic diagram of the side-incident dark-field setup. (d1) Side-incident dark-field spectra and (d2) CD spectra of Au/CNC with circularly-polarized longitudinal and transverse side-incident light. Side-incident dark-field spectra of Au/CNC with different circularly-polarized (e1) longitudinal and (f1) transverse side-incident light at different polarization analyzer angles. CD spectra obtained from the spectra in (e2) Figure e1 and (f2) Figure f1. The insets at the right bottom right corner of each figure indicate the illumination approach.

## 2D dispersive characteristic of the Au/CNC

Along with the different optical response, the Au/CNC presents significant different color distributions at different incident projective angles. The side-incident microscopic images collected by an objective of 50×, NA = 0.33 are shown in the top row of Figure 3. With the increase of the incident projective angle from 0° to 90°, the single Au/CNC successively turns to red, yellow, white, cyan and orange with a spatial distribution. This irregularly color variation is amazing. To reveal the origin of the colors, we analyzed the emission direction of the scattered light by imaging the Au/CNC on Fourier plane. In Fourier plane images of Figure 3, the direction distribution of the emission light (collected by an objective of 100×, NA = 0.9) was transformed to the spatial distribution. The further a light dot locates from the center in Fourier plane image, the larger emission angle at that direction for the light. It is clear that the Au/CNC disperses the incident light into colorful bands perpendicular to the axis of the coil like a grating. At an incident projective angle of 90°, the zero-order diffraction band locates in the middle of the image and the first-order diffraction bands locate symmetrically on its sides. As the angle decreases, the bands shift along the incident direction, and the second-order diffraction band appears. The zero order diffraction band concentrates most of the diffraction light, and it is the reason why the intensity of the spectra for longitudinal incident condition in Figure 2d, e, f is much larger than that for transverse incident condition (without zero order diffraction band). This dispersive effect demonstrates the Au/CNC can be employed as an effective dispersive component for a microspectrometer.

The dotted circles in Fourier plane images indicate the region of NA = 0.33, at which the real space images of the Au/CNC on the top row were collected. In this region, the whole width of the zero- or first-order diffraction bands were included at the incident projective angle of 30°-60°, 80°-90°, and thus the Au/CNC shows bright white or orange appearance. At other incident projective angles, only a part of diffraction bands in width is in the region, so the Au/CNC presents the included colors with spatial distribution.

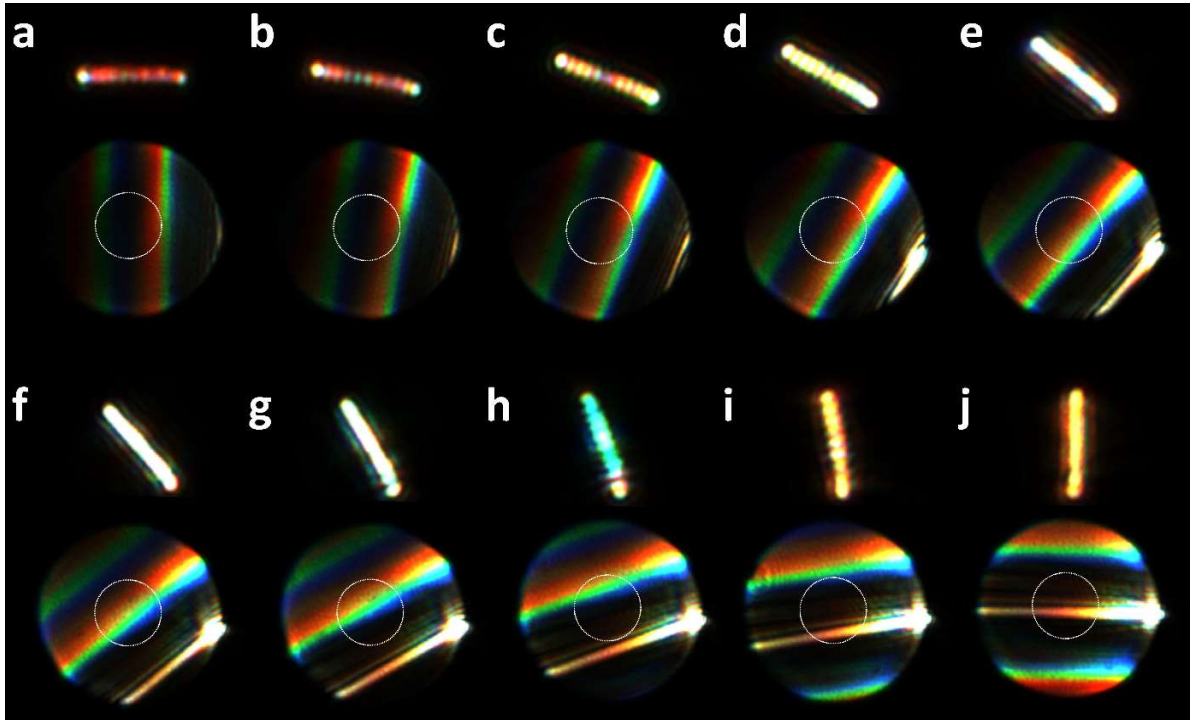

**Figure 3.** Side-incident microscopic images of the Au/CNC with an objective of 50×, NA = 0.33 (top row) and corresponding Fourier plane images with an objective of 100×, NA = 0.9 (bottom row) at incident projective angle of (a) 0°, (b) 10°, (c) 20°, (d) 30°, (e) 40°, (f) 50°, (g) 60°, (h) 70°, (i) 80°, and (j) 90° to the axis of the coil. The dotted circles in Fourier plane images correspond to the region of NA = 0.33, at which the top images were collected.

For a grating-based spectrometer, the resolution is proportional to the diffraction order of the band used. Accordingly, we investigated the dispersive properties at the incident projective angle of 0°, at which both the first- and second-order diffraction bands with relatively uniform brightness distribution are well included in Fourier plane image. Incident light with different polarization conditions, i.e. non-polarized, LCP, RCP, and linearly-polarized with different angles, were employed, and the corresponding Fourier plane images are shown in Figure S1. It can be seen that the brightness distribution of the bands is quite different for the different polarized incident light. This difference is magnified by introducing the polarization analyzer with an angle of 90° (Figure 4), which also gives rise to enhanced CD in Figure 2e2, f2. Due to the chiral structure of the Au/CNC, the brightness distribution for LCP and RCP incident light is significantly different in Figure 4b, c, and their bright and dark areas of the bands are almost complementary. For the linearly-polarized incident light in Figure 4d-i, the position and length of the bright area regularly changes with the incident polarization angle for both bands. As the angle increases from 0° to 60°, the main bright area extends upward. When the angle further increases from 60° to 180° (0°), the main bright area shortens and finally turns into the bottom bright area.

Benefiting from the circular curved outer surface, the Au/CNC scatters light in all the directions perpendicular to the axis of the coil, hence the incident light is dispersed into bands instead of dots. At different incident polarization conditions, the incident light is scattered from different areas of the outer surface of the Au/CNC into different directions, thus exhibits bands with different bright areas. This distinctive dispersive effect enables Au/CNC to distinguish the polarization condition as well as exhibit the wavelength distribution of the incident light. So, a novel 2D wavelength-polarization microspectrometer can be obtained based on the Au/CNC dispersive component.

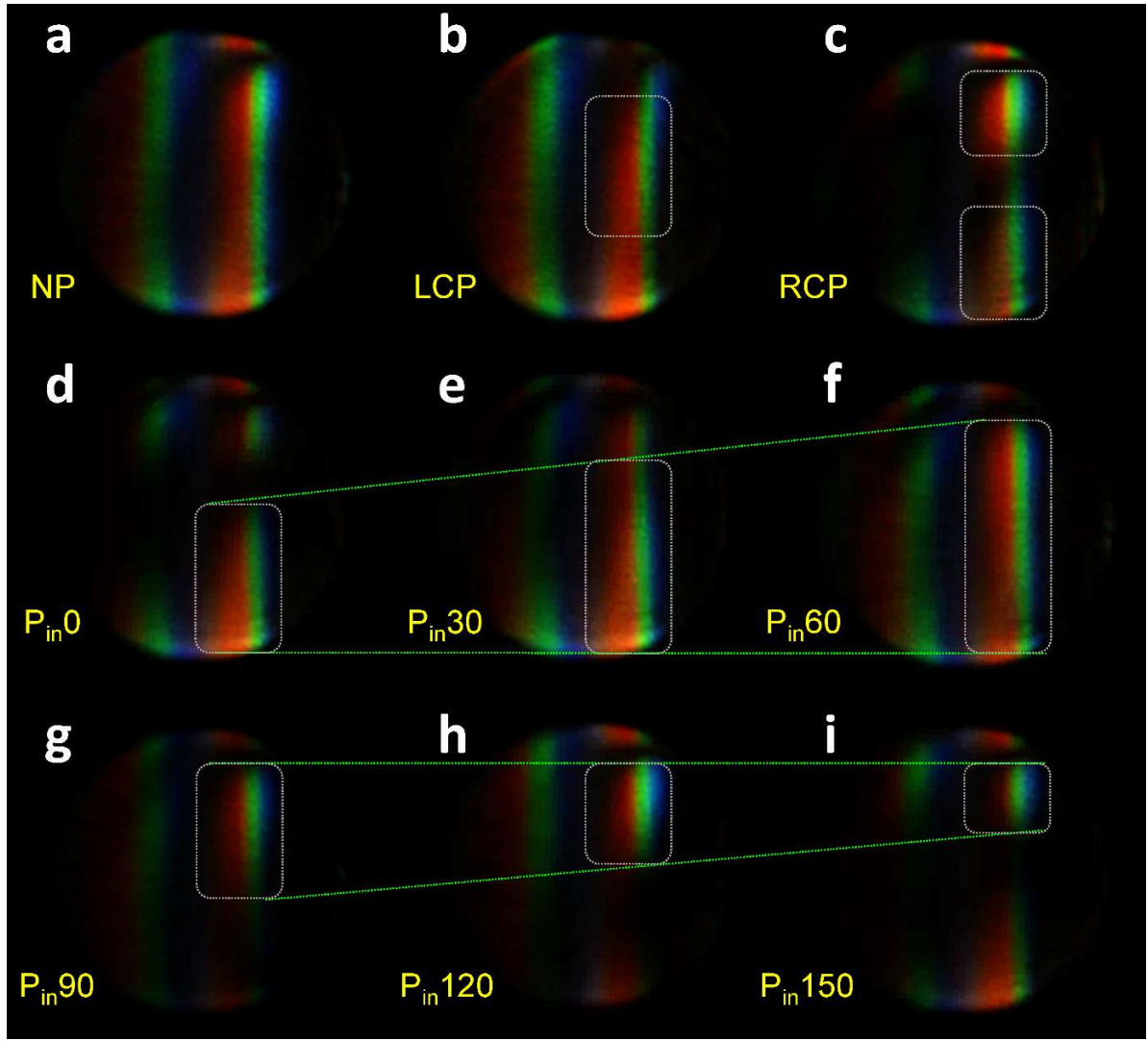

**Figure 4.** Fourier plane images of Au/CNC with (a) non-polarized, (b) LCP, (c) RCP, and (d) 0°, (e) 30°, (f) 60°, (g) 90°, (h) 120°, (i) 150° linear polarized incident light at polarization analyzer angle of 90°. The gray dotted rounded rectangles indicate the main bright areas of the first-order diffraction band, and the green dotted lines indicate the variation trend of the bright areas with linear polarization angles.

**Monochromatic analysis**

In practical application of spectrometers, light signals are usually collected by monochrome detector instead of color detector, and the wavelength is known by the position of signal lines. Here, the dispersive ability of the Au/CNC was also demonstrated with a monochrome CCD, and the incident light is the combination of 550 and 650 nm monochromatic light. In Figure 5a-i, four straight lines corresponding to the wavelength of 550 and 650 nm in the first- and second-order bands can be clearly distinguished. The position of the lines is not influenced by the incident polarization condition, so that the wavelength of the incident light can be identified by the location of the lines in the images, as the spectrum with the pixel in Figure 5j shown.

The y-axis intensity distributions of the line corresponding to 650 nm in the first-order diffraction band for each incident polarization condition were exhibited in the right panels of Figure 5a-i. Consistent with color Fourier plane images in Figure 4, these intensity distribution spectra have close relation with their incident polarization conditions. To eliminate the interference of inherent optical response of the Au/CNC, the intensity distribution spectra for LCP and RCP incident light were divided by that for non-polarized incident light and compared in Figure 5k. Due to the great chiral structure of the Au/CNC, the profiles of the spectra are well opposite in most regions. This means LCP and RCP light can be easily distinguished by the microspectroscope with the Au/CNC dispersive component.

For linearly-polarized incident conditions, the intensity and position of the peaks in the intensity distribution spectra vary regularly with the polarization angle, and the variation of the first- and second-order diffraction bands (Figure S2) is not the same. To indicate the incident polarization angle, we calculated the y-axis intensity ratio of the 134th pixel in the second-order diffraction band and the 85th pixel in the first-order diffraction band of the 650 nm lines. As shown in Figure

5l, the intensity ratio ($R$) has an e-exponential relation with the incident linear polarization angle of

$$R = 0.12e^{\frac{P_{in}}{89}} + 0.16 \tag{1}$$

where $P_{in}$ is the incident linear polarization angle in degree. Based on the equation above, the incident linear polarization angle can be obtained with a quite high accuracy. These demonstrate that the anisotropic scattering of the Au/CNC can effectively measure the polarization condition of the incident light, and it also verifies the feasibility of the 2D wavelength-polarization microspectroscope.

Dispersive performance of the bare CNC was also shown in Figure S3. Besides the weaker scattering ability of the CNC as its spectra shown in the above optical response section, the signal lines dispersed by CNC are twisted. This make it hard to give accurate wavelength and polarization information. Therefore, the supported Au NPs on the CNC is essential for the dispersive component in the microspectroscope. They not only enhance the scattering ability by LSPR, but also greatly improve the quality of the dispersed signal lines via modifying the scattering direction of the helical structure.

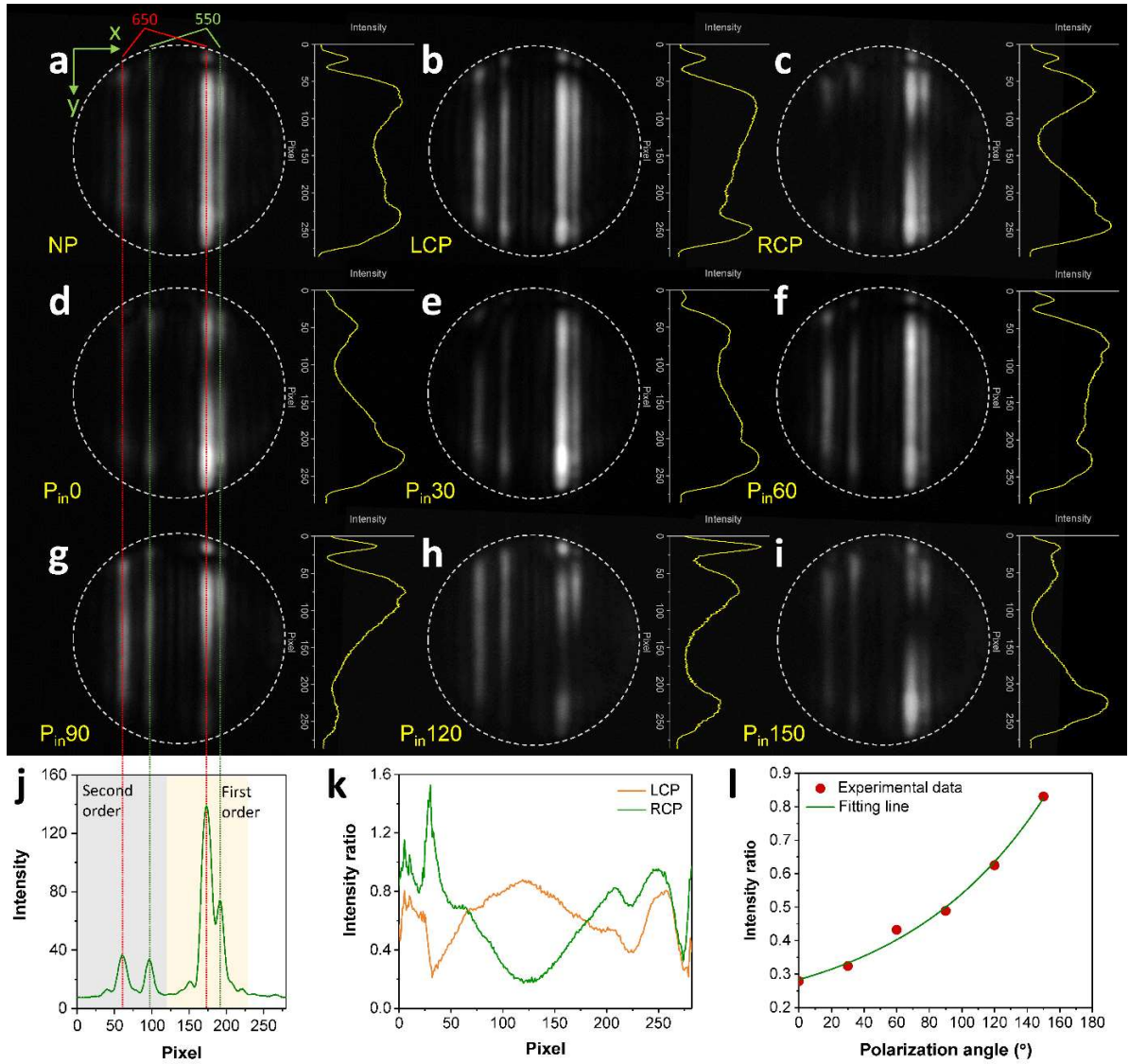

**Figure 5.** Monochromatic Fourier plane images (left panel) and corresponding y-axis intensity distribution of the 650 nm line in the first-order diffraction band (right panel) for the Au/CNC with (a) non-polarized, (b) LCP, (c) RCP, and (d) 0°, (e) 30°, (f) 60°, (g) 90°, (h) 120°, (i) 150° linear polarized 550 and 650 nm incident light at polarization analyzer angle of 90°. The dotted circles indicate the imaging area limited by the objective NA. (j) Average intensity distribution in x axis of Fourier plane image for the Au/CNC in Figure a. (k) The ratio of the intensity in Figure b and c with that in Figure a. (i) The y-axis intensity ratio of the 134th pixel in the second-order

diffraction band and the 85th pixel in the first-order diffraction band of the 650 nm lines with the linear polarization angle of the incident light.

**Performance parameters for a microspectrometer**

Resolution power and measurable band width are important metrics to evaluate the performance of a spectrometer. These parameters for the Au/CNC dispersive component are measured by changing the incident wavelength. As shown in Figure S4a, the lines of 550 and 630 nm incident light can be distinguished in the first-order diffraction band, indicating the resolution at least 80 nm. For the second-order band, even 550 and 590 nm can be directly distinguished in Figure S4b with resolution exceeding 40 nm. This twice magnitude of the resolution power for the two order bands is consistent with grating-based spectrometer resolution equation of

$$R = kN \qquad (2)$$

where $R$, $k$, $N$ are resolution power, diffraction order and the number of grating grooves, respectively. At this resolution, incident light of four different wavelengths with an equal distance of 50 nm can be clearly distinguished in Figure S4c. And the four corresponding signal lines are equidistant, indicating the good linear relation of wavelength and the signal line position.

In Figure S4d, the lines of 500 nm can be seen in both the first- and second-order diffraction bands, even though the brightness is quite low, thus the minimum measurable wavelength is about 500 nm. For the maximum measurable wavelength, it is as high as 900 nm in the near-infrared range for the first-order diffraction band (Figure S4e), and that for the second-order diffraction band is about 750 nm (Figure S4f), limited by the NA of the objective. It is noted that the quantum efficiency of the CCD decreases quickly with the wavelength in the near-infrared range, and it is just below 20% at 900 nm. This implies the maximum measurable wavelength may exceed 900 nm. Therefore, the measurable band width for the Au/CNC dispersive component is evaluated to be at least 400 and 250 nm for the first- and second-order diffractions.

Here, the 40-80 nm resolution power of the Au/CNC dispersive component is not quite high compared with other works [1]. However, it is far below the resolution limit for the Au/CNC. The period of the Au/CNC in this work is only 11. According to Equation 1, if a longer Au/CNC with more periods was used, the resolution will remarkably increase. Meanwhile, the larger scattering cross section will also improve the sensitivity. On the other hand, computer processing is an important part in modern measurements. It significantly improves the signal-to-noise ratio and the resolution of obtained signals. Employing computer processing, it will further achieve much higher resolution power of the Au/CNC dispersive component. It is also noted that a polarizer was used in the optical path to improve the polarization sensitivity of the spectroscope. Nevertheless, it will not enlarge the footprint in a practical microspetrometer by depositing a polarizer film on the detector, for the polarizer orientation is not tuned during application. Considering this, the quite large measurable band width (≥400 nm) and ultra-small size make the Au/CNC an outstanding candidate for the microspectrometer. And even more important, the Au/CNC achieve the wavelength dispersion, as well as the polarimetric ability, including chiral light distinguishing and polarization orientation measurement. This produces an extremely powerful 2D wavelength-polarization microspectroscope.

## Conclusions

In summary, we prepared the hybrid plasmonic helical Au/CNC and explored its optical response and dispersive property. The coil has an anisotropic optical response, and the chiral structure endues it significantly different response to RCP and LCP light. LSPR of the Au NPs significantly enhances the scattering ability and modifies the scattering direction of the structure. Fourier plane images exhibit a unique 2D wavelength-polarization dispersive characteristic of the Au/CNC. Along the axis of the coil, the incident light is dispersed as wavelength to bands of different diffraction orders like a grating. A resolution power of 40-80 nm and quite large measurable band width at least 400 nm can be obtained in the visible-near-infrared range. While in the direction perpendicular to the axis of the coil, incident light is dispersed as polarization with bright and dark areas. We are able to not only distinguish LCP and RCP light, but also obtain the polarization orientation of linearly-polarized light directly. This wonderful dispersion effect makes Au/CNC work as a dispersive component for a 2D wavelength-polarization microspectrometer, which can simultaneously measure the wavelength and polarization condition of the light. This technique will revolutionize the experimental researching approaches on polarization, realize the polarization analysis of dynamic objects. Moreover, there will be new revolutionary techniques developed via integrating the microspectrometer with other systems.

## Methods

**Synthesis of Au/CNC**

The helical CNC was prepared by CVD on the $\alpha$-$Fe_2O_3$/$SnO_2$ catalyst with acetylene as the precursor. The catalyst was obtained from soluble $Fe^{3+}$ and $Sn^{4+}$ salts via hydrothermal reaction at 180 °C, followed by spin coating on a Si substrate. The growth of CNC was performed at 710 °C with a mixture of 235 sccm Ar and 25 sccm acetylene flowing over the catalyst. Detailed procedures can be referred to the reported literature [30-31]. The as-prepared CNC was scrapped from the substrate and dispersed in 20 mL of water under ultrasonication. For the direct growth of Au NPs on the CNC, 1 mL of the CNC solution was added to 19 mL of water in a round-bottomed flask and stirred at 90°C. Then, 0.3 mL of sodium citrate solution (0.34 M) and 60 μL of HAuCl$_4$ solution (0.1 M) were sequentially injected (time delay 5 min). The solution was refluxed for 15 min, and was then cooled to room temperature under continuous stirring. The product was washed with water for three times and finally redispersed in ethanol for characterization.

**Characterization and optical measurements**

The HAADF-STEM images and EDX elemental mapping were obtained with FEI Titan cubed Themis G2 300 microscope equipped with a probe aberration corrector and monochromator operating at 200 kV. The high resolution morphology and crystal features of the Au/CNC were analyzed by TEM (JEOL JEM-2100F) at accelerating voltage of 200 kV. The Au/CNC specimen on silicon substrate for optical measurements was observed by SEM (TESCAN MIRA3) at accelerating voltage of 20 kV with second electron (SE) and backscattered electron (BSE) detectors, respectively. The specimen was prepared by dropping dilute ethanol solution of the as-prepared Au/CNC on a silicon substrate followed by drying.

The identical single isolated Au/CNC for SEM characterization was employed to perform the optical measurements with an optical microscope (Olympus BX53) working in reflection dark-field mode. Spectra of the specimen were collected by a spectrometer (BWTEK Exemplar plus) via dry objectives of 50×, NA = 0.5 or 100×, NA = 0.9, illuminated with the build-in halogen lamp or side-incident beam from the supercontinuum source filtered by an acousto-optic tunable filter (AOTF), respectively. Fourier plane and real space images of the Au/CNC under side-incident illumination were collected by the setup as shown in Figure S5 with a color CCD (TUCSEN TCC-1.4HICE) or a monochrome CCD (Qimaging Retiga R1). Air objectives of 100×, NA = 0.9 and 50×, NA = 0.33 (with a homemade iris) were used.

## Declaration of Competing Interest

The authors declare that they have no known competing financial interests or personal relationships that could have appeared to influence the work reported in this paper.

## Acknowledgements

This research was supported by the National Natural Science Foundation of China (Grant No. 12074054, 11704058, 51972039) and the Fundamental Research Funds for the Central Universities (Grant No. DUT21LK06).

## References

[1]     Z. Yang, T. Albrow-Owen, W. Cai, T. Hasan, Miniaturization of optical spectrometers. Science 371 (2021) eabe0722.

[2]     R. A. Crocombe, Portable Spectroscopy. Appl. Spectrosc. 72 (2018) 1701-1751.

[3]     A. J. S. McGonigle, T. C. Wilkes, T. D. Pering, J. R. Willmott, J. M. Cook, F. M. Mims, A. V. Parisi, Smartphone Spectrometers. Sensors 18 (2018) 223.

[4]     J. Wang, B. Zheng, X. Wang, Strategies for high performance and scalable on-chip spectrometers. J. Phys.-Photon. 3 (2020) 012006.

[5]     M. Faraji-Dana, E. Arbabi, A. Arbabi, S. M. Kamali, H. Kwon, A. Faraon, Compact folded metasurface spectrometer. Nat. Commun. 9 (2018) 4196.

[6]     R. Cheng, C. L. Zou, X. Guo, S. Wang, X. Han, H. X. Tang, Broadband on-chip single-photon spectrometer. Nat. Commun. 10 (2019) 4104.

[7]     A. Y. Zhu, W.-T. Chen, M. Khorasaninejad, J. Oh, A. Zaidi, I. Mishra, R. C. Devlin, F. Capasso, Ultra-compact visible chiral spectrometer with meta-lenses. APL Photon. 2 (2017) 036103.

[8]     A. Tittl, A. Leitis, M. K. Liu, F. Yesilkoy, D. Y. Choi, D. N. Neshev, Y. S. Kivshar, H. Altug, Imaging-based molecular barcoding with pixelated dielectric metasurfaces. Science 360 (2018) 1105-1109.

[9]     J. Bao, M. G. Bawendi, A colloidal quantum dot spectrometer. Nature 523 (2015) 67-70.

[10]    X. Zhu, L. Bian, H. Fu, L. Wang, B. Zou, Q. Dai, J. Zhang, H. Zhong, Broadband perovskite quantum dot spectrometer beyond human visual resolution. Light Sci. Appl. 9 (2020) 73.

[11]    D. Pohl, M. R. Escale, M. Madi, F. Kaufmann, P. Brotzer, A. Sergeyev, B. Guldimann, P. Giaccari, E. Alberti, U. Meier, R. Grange, An integrated broadband spectrometer on thin-film lithium niobate. Nat. Photonics 14 (2020) 24-29.

[12]    E. Le Coarer, S. Blaize, P. Benech, I. Stefanon, A. Morand, G. Lerondel, G. Leblond, P. Kern, J. M. Fedeli, P. Royer, Wavelength-scale stationary-wave integrated Fourier-transform spectrometry. Nat. Photonics 1 (2007) 473-478.

[13]    S. N. Zheng, J. Zou, H. Cai, J. F. Song, L. K. Chin, P. Y. Liu, Z. P. Lin, D. L. Kwong, A. Q. Liu, Microring resonator-assisted Fourier transform spectrometer with enhanced resolution and large bandwidth in single chip solution. Nat. Commun. 10 (2019) 2349.

[14]    Z. Y. Yang, T. Albrow-Owen, H. X. Cui, J. Alexander-Webber, F. X. Gu, X. M. Wang, T. C. Wu, M. H. Zhuge, C. Williams, P. Wang, A. V. Zayats, W. W. Cai, L. Dais, S. Hofmann, M. Overend, L. M. Tong, Q. Yang, Z. P. Sun, T. Hasan, Single-nanowire spectrometers. Science 365 (2019) 1017-1020.

[15]  B. Redding, S. F. Liew, R. Sarma, H. Cao, Compact spectrometer based on a disordered photonic chip. Nat. Photonics 7 (2013) 746-751.

[16]  W. Hartmann, P. Varytis, H. Gehring, N. Walter, F. Beutel, K. Busch, W. Pernice, Waveguide‐integrated broadband spectrometer based on tailored disorder. Adv. Opt. Mater. 8 (2020) 1901602.

[17]  R. F. Wolffenbuttel, State-of-the-Art in Integrated Optical Microspectrometers. IEEE T. Instrum. Meas. 53 (2004) 197-202.

[18]  Z. Y. Li, M. J. Deen, Q. Y. Fang, P. R. Selvaganapathy, Design of a flat field concave-grating-based micro-Raman spectrometer for environmental applications. Appl. Optics 51 (2012) 6855-6863.

[19]  C. Sciancalepore, R. J. Lycett, J. A. Dallery, S. Pauliac, K. Hassan, J. Harduin, H. Duprez, U. Weidenmueller, D. F. G. Gallagher, S. Menezo, B. B. Bakir, Low-Crosstalk Fabrication-Insensitive Echelle Grating Demultiplexers on Silicon-on-Insulator. IEEE Photon. Technol. Lett. 27 (2015) 494-497.

[20]  T. A. Kwa, R. F. Wolffenbuttel, Integrated grating detector array fabricated in silicon using micromachining techniques. Sens. Actuators A 31 (1992) 259-266.

[21]  N. Ghosh, I. A. Vitkin, Tissue polarimetry: concepts, challenges, applications, and outlook. J. Biomed. Opt. 16 (2011) 110801.

[22]  S. Trippe, Polarization And Polarimetry: A Review. J. Korean Astron. Soc. 47 (2014) 15-39.

[23]  J. S. Tyo, D. L. Goldstein, D. B. Chenault, J. A. Shaw, Review of passive imaging polarimetry for remote sensing applications. Appl. Optics 45 (2006) 5453-5469.

[24] X. Wang, Y. Wang, W. Gao, L. Song, C. Ran, Y. Chen, W. Huang, Polarization-Sensitive Halide Perovskites for Polarized Luminescence and Detection: Recent Advances and Perspectives. Adv. Mater. 33 (2021) e2003615.

[25] H. Raghubanshi, E. D. Dikio, E. B. Naidoo, The properties and applications of helical carbon fibers and related materials: A review. J. Ind. Eng. Chem. 44 (2016) 23-42.

[26] D.-C. Wang, Y. Lei, W. Jiao, Y.-F. Liu, C.-H. Mu, X. Jian, A review of helical carbon materials structure, synthesis and applications. Rare Metals 40 (2020) 3-19.

[27] Y. Sun, C. Wang, L. Pan, X. Fu, P. Yin, H. Zou, Electrical conductivity of single polycrystalline-amorphous carbon nanocoils. Carbon 98 (2016) 285-290.

[28] L. Li, L. Lu, S. Qi, Preparation, characterization and microwave absorption properties of porous nickel ferrite hollow nanospheres/helical carbon nanotubes/polypyrrole nanowires composites. J. Mater. Sci.-Mater. El. 29 (2018) 8513-8522.

[29] J. A. Schuller, E. S. Barnard, W. S. Cai, Y. C. Jun, J. S. White, M. L. Brongersma, Plasmonics for extreme light concentration and manipulation. Nat. Mater. 9 (2010) 193-204.

[30] Y. Zhao, J. Wang, H. Huang, T. Cong, S. Yang, H. Chen, J. Qin, M. Usman, Z. Fan, L. Pan, Growth of carbon nanocoils by porous alpha-Fe2O3/SnO2 catalyst and Its buckypaper for high efficient adsorption. Nano-Micro Lett. 12 (2020) 23.

[31] Y. Zhao, J. Wang, H. Huang, H. Zhang, T. Cong, D. Zhang, N. Wen, Y. Zhang, Z. Fan, L. Pan, Catalytic anisotropy induced by multi-particles for growth of carbon nanocoils. Carbon 166 (2020) 101-112.

# 2D wavelength-polarization dispersive microspectroscope based on a hybrid plasmonic helical nanostructure

*Zhiguang Sun*[a]*, Huan Chen*[b]*, Zhenglong Zhang*[b]*, Lujun Pan*[a]*, Yiming Yang*[c]*, Bin Dong*[d,*] *and Yurui Fang*[a,*]

[a] Key Laboratory of Materials Modification by Laser, Electron, and Ion Beams (Ministry of Education); School of Physics, Dalian University of Technology, Dalian 116024, P.R. China

[b] School of Physics and Information Technology, Shaanxi Normal University, Xi'an 710062, P. R. China

[c] School of Microelectronics, Dalian University of Technology, Dalian 116024, P.R. China

[d] School of Physics and Materials Engineering, Dalian Minzu University, Dalian 116600, P. R. China

[*] Corresponding authors: yrfang@dlut.edu.cn (Fang); dong@dnu.edu.cn (Dong)

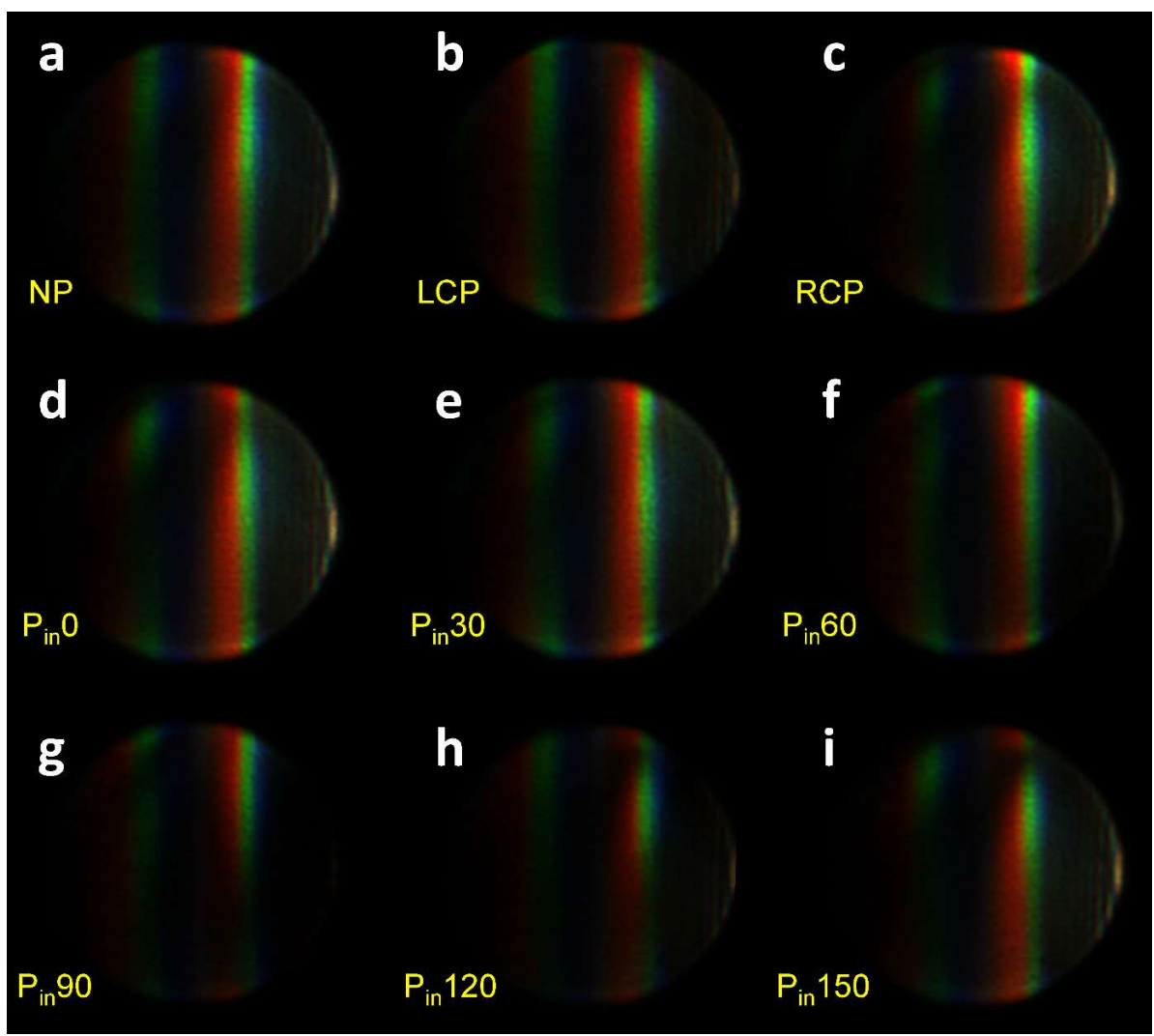

**Figure S1.** Fourier plane images of Au/CNC with (a) non-polarized, (b) LCP, (c) RCP, and (d) 0°, (e) 30°, (f) 60°, (g) 90°, (h) 120°, (i) 150° linear polarized incident light.

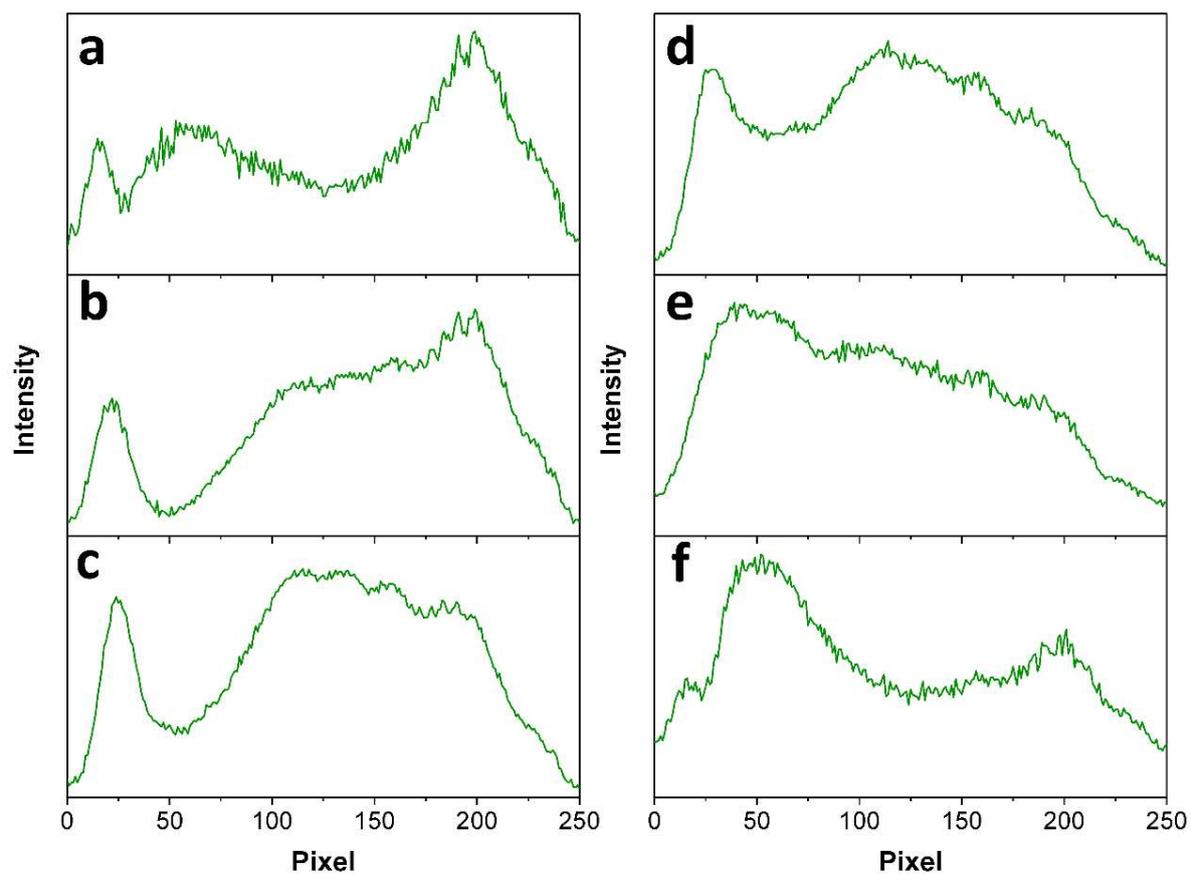

**Figure S2.** Y-axis intensity distribution of the 650 nm line in the second-order diffraction band for the Au/CNC with (a) 0°, (b) 30°, (c) 60°, (d) 90°, (e) 120°, (f) 150° linear polarized 550 and 650 nm incident light at polarization analyzer angle of 90°.

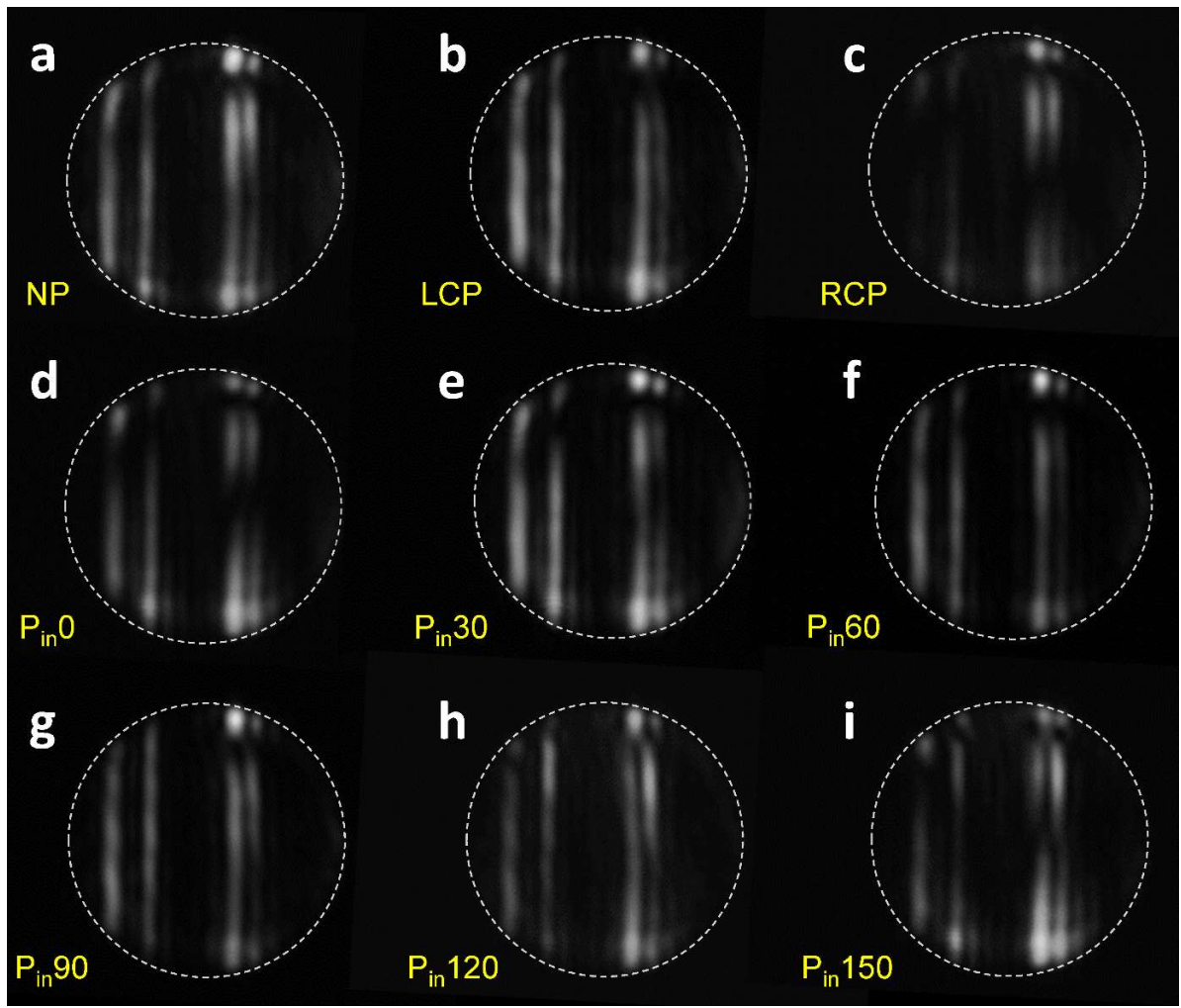

**Figure S3.** Monochromatic Fourier plane images of CNC with (a) non-polarized, (b) LCP, (c) RCP, and (d) 0°, (e) 30°, (f) 60°, (g) 90°, (h) 120°, (i) 150° linear polarized 550 and 650 nm incident light at polarization analyzer angle of 90°. The dotted gray circles in Fourier plane images indicate the imaging area limited by the NA of the objective.

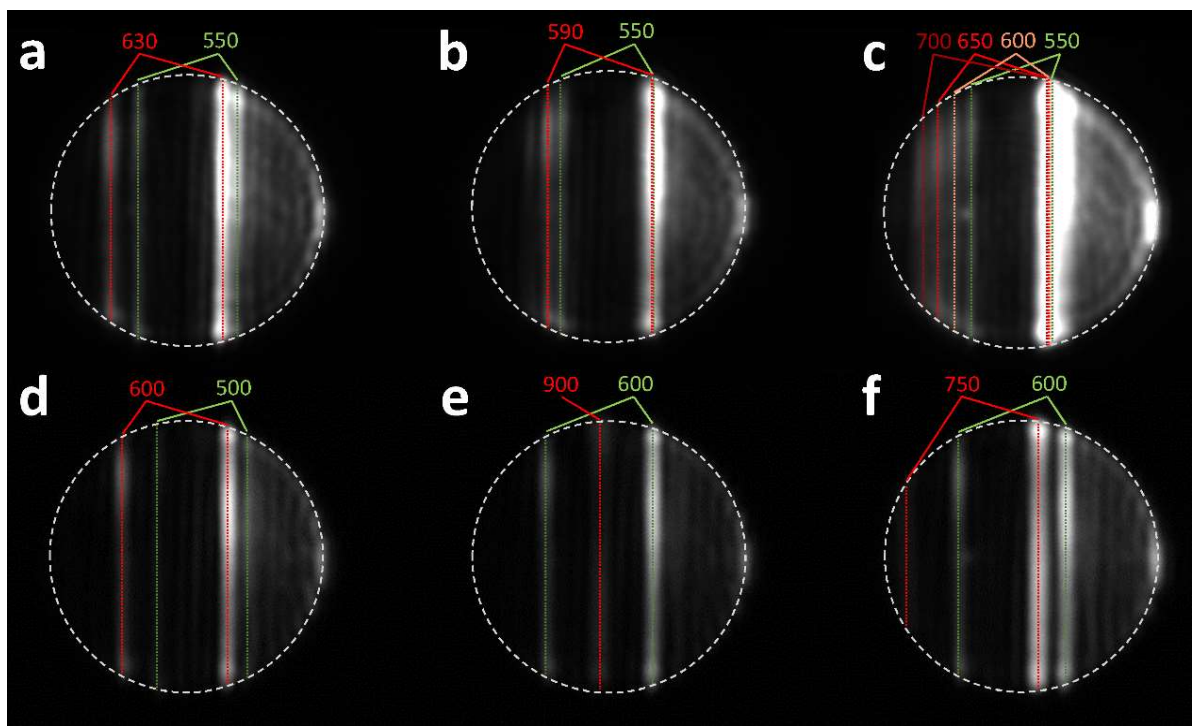

**Figure S4.** Monochromatic Fourier plane images for the Au/CNC with non-polarized (a) 550, 630 nm, (b) 550, 590 nm, (c) 550, 600, 650, 700 nm, (d) 500, 600 nm, (e) 600, 750 nm, and (f) 600, 900 nm incident light at polarization analyzer angle of 90°. The dotted gray circles in Fourier plane images indicate the imaging area limited by the NA of the objective.

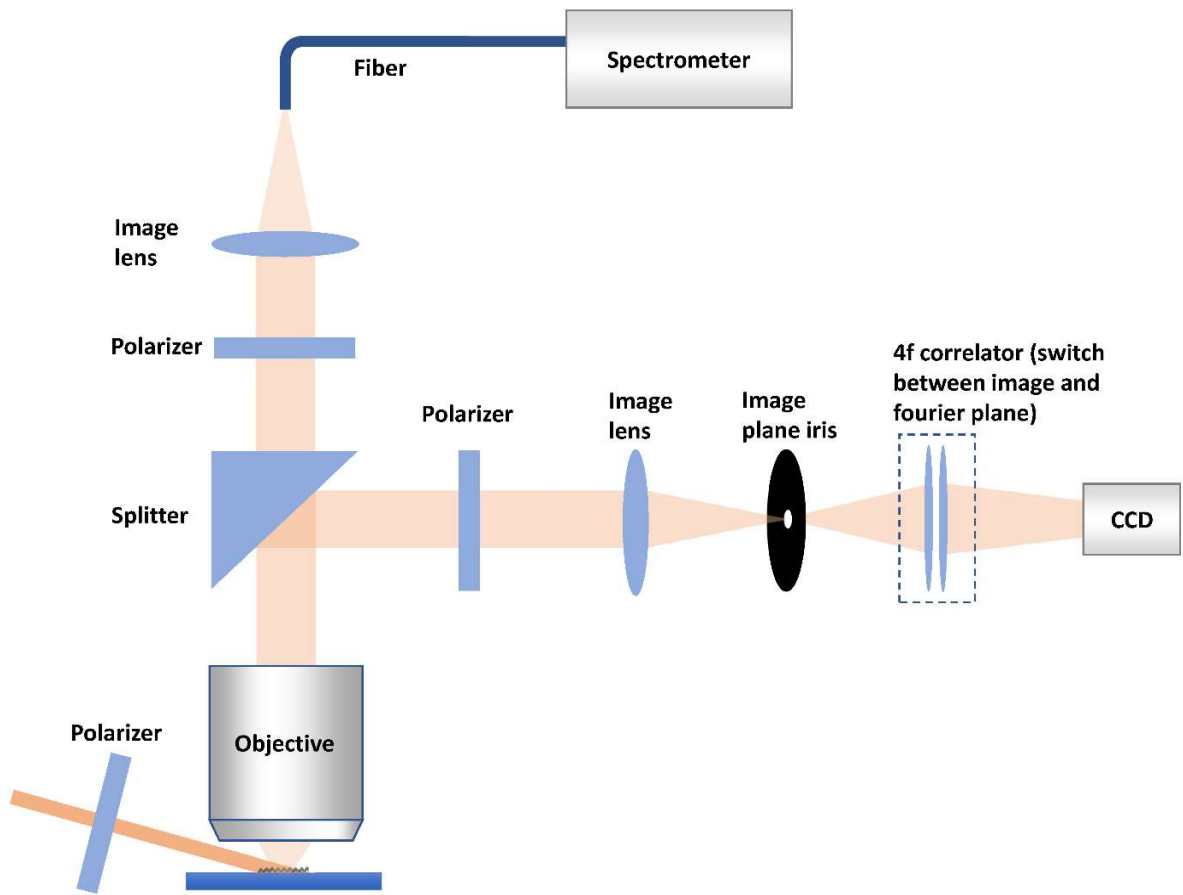

**Figure S5.** Setup for spectra and Fourier plane images collection.